\documentclass[twocolumn,english,amsart,showpacs,preprintnumbers,amsmath,amssymb,floatfix]{revtex4-1}
\usepackage{tikz,xcolor}
\usepackage[colorlinks = true,
linkcolor = blue,
urlcolor  = blue,
citecolor = blue,
anchorcolor = blue]{hyperref}

\definecolor{lime}{HTML}{A6CE39}
\DeclareRobustCommand{\orcidicon}{%
	\begin{tikzpicture}
		\draw[lime, fill=lime] (0,0)
		circle [radius=0.16]
		node[white] {{\fontfamily{qag}\selectfont \tiny ID}};
		\draw[white, fill=white] (-0.0625,0.095)
		circle [radius=0.007];
	\end{tikzpicture}
	\hspace{-2mm}
}

\foreach \x in {A, ..., Z}{%
	\expandafter\xdef\csname orcid\x\endcsname{\noexpand\href{https://orcid.org/\csname orcidauthor\x\endcsname}{\noexpand\orcidicon}}
}

\usepackage[T1]{fontenc}
\usepackage[latin9]{inputenc}
\usepackage{color}
\usepackage{array}
\usepackage{amstext}
\usepackage{graphicx}
\usepackage{esint}
\usepackage{rotating}
\usepackage{appendix}
\usepackage{float}

\usepackage[font=small,labelfont=bf]{caption}
\usepackage{xcolor}
\usepackage{ulem}

\makeatletter

\newcommand{\tA}{{\widetilde {\mathcal{A}}}}

\newcommand{\tk}{{\widetilde k}}

\newcommand\RV{\mbox{\boldmath $R$}}
\newcommand\eV{\mbox{\boldmath $e$}}
\newcommand\IV{\,\mbox{\boldmath $I$}}
\newcommand\PV{\mbox{\boldmath $P$}}                           
\def\as{\relax\ifmmode \alpha_s\else{$ \alpha_s${ }}\fi}  
\def\abar{\relax\ifmmode{\bar{a}}\else{$\bar{a}${ }}\fi}  
\def\be{\begin{equation}}
\def\ee{\end{equation}}
\def\ba{\begin{eqnarray}}
\def\ea{\end{eqnarray}}

\makeatletter

\@ifundefined{textcolor}{}
{%
 \definecolor{BLACK}{gray}{0}
 \definecolor{WHITE}{gray}{1}
 \definecolor{RED}{rgb}{1,0,0}
 \definecolor{GREEN}{rgb}{0,1,0}
 \definecolor{BLUE}{rgb}{0,0,1}
 \definecolor{CYAN}{cmyk}{1,0,0,0}
 \definecolor{MAGENTA}{cmyk}{0,1,0,0}
 \definecolor{YELLOW}{cmyk}{0,0,1,0}
 }

\@ifundefined{definecolor}
 {\usepackage{color}}{}
\@ifundefined{definecolor}
 {\usepackage{color}}{}
\makeatother
\usepackage{babel}

\begin{document}


\title {Analytical perturbation theory and Nucleon structure function in infrared region}

\author { L.~Ghasemzadeh$^{1}$}
\email{leilaghasemzadeh@stu.yazd.ac.ir}

\author{ A.~Mirjalili$^{1}$\orcidA{}}
\email{a.mirjalili@yazd.ac.ir (corresponding author)}

\author {S.~Atashbar Tehrani$^{2}$\orcidB{}}
\email{atashbar@ipm.ir}

\affiliation {
$^{(1)}$Physics Department, Yazd University, P.O.Box 89195-741, Yazd, Iran       \\
$^{(2)}$School of Particles and Accelerators, Institute for Research in Fundamental Sciences (IPM), P.O.Box
19395-5531, Tehran, Iran}

\date{\today}

%
\begin{abstract}\label{abstract}
We employ {{} analytic QCD (anQCD) approach to analyze} the unpolarized nucleon structure function (NSF) in  deep inelastic scattering ( DIS )  processes  at the next-to-leading order  (NLO) accuracy. Considering the unreliable results of underlying perturbative QCD (pQCD)  at energy scale  $Q^2\sim\Lambda ^2$ and even less we modify the calculations at these scales using anQCD approach and compare them with results from underlying pQCD and also with the available experimental data. In these progresses  the massive perturbation theory (MPT) model is also  used where an effective mass  is attributed to gluons.
We finally  use the Jacobi polynomials formalism to transfer the calculations from Mellin moment space to Bjorken-$x$ space. {{}  To confirm the validity of anQCD approach the Gottfired sum rule is also investigated. The achieved numerical results at low energy scales {{} are compatible with} what is expected and corresponding to an admissible behaviour of parton densities}.

\end{abstract}

\maketitle


%
\section{Introduction}\label{Introduction}
 An observable should be an analytic (holomorphic) function {{}in the complex $Q^2$ plane where $Q^2 \not< 0$.}
 At high  energy scales, i.e. {{}$|Q^2| \gg 1$,}  we can have a good theoretical description and achieve reliable results which are confirming experimental data, using underlying pQCD. But at the energy scale {{} near to QCD cutoff parameter, i.e.,} $Q^2 \sim \Lambda ^2$ and less, the coupling constant of QCD starts to be growing rapidly and {{} as a result, facing Landau IR-singularities}. {{}On the other hand} {{} the spacelike QCD observable, such as the nucleon structure functions, do not have such singularities.}
Accordingly one cannot obtain any reliable results from underlying pQCD, thence we need an efficient approach {{} which eliminates these singularities} in order to achieve suitable results.
There are various approaches to attain this goal such as Brodsky coupling constant achievement by Ads/CFT \cite{Brodsky:2010ur}, the dispersive approach of Dokshitzer \cite{Dokshitzer:1995qm,Dokshitzer:1995zt} and finally Analytic Perturbation Theory(APT)\cite{ShS,MS,MSS,MSa,Sh1,Sh2}.
We use the last one to shift  and {{} even eliminates the mentioned singularities in calculations} of physical quantities such as unpolarized nucleon structure function (NSF) and also the Gottfried sum rule, {{} and thus modify} their theoretical predictions. {{} We refer to \cite{Ours} for recent related work.}
In this approach, the running QCD coupling constant {{}($a(Q^2) = \frac{\alpha_s(Q^2)}{\pi}$)} is {{} transformed in an analytic function of $Q^2$ (analytic for $Q^2 \not<0$) which is called analytic QCD coupling constant [$A_1(Q^2)$] that does not have any Landau singularities and} we are able to calculate {{} the results} for the quantities {{} without any such singularities at the low energies, using} the analytic coupling constant.

{{}Among the approaches which eliminate the Landau singularities, we can refer to Fractional Analytic Perturbation Theory [FAPT] \cite{BMS1,BMS2,BMS3,Ayala:2014pha}; $2 \delta$anQCD \cite{2danQCD,Ayala:2014pha} and $3\delta$anQCD \cite{3danQCD} which are based on parameterising the spectral function at low energies by two or three Dirac delta functions, respectively; and finally massive perturbation theory (MPT) \cite{MPT,Ayala:2014pha} which is based on removing the Landau singularities by shifting them into the timelike region.}
As an attribute that the last method possesses, it considers an effective mass for the gluon. Since we are working on nucleon structure function which contains  the singlet and gluon sectors , we decide to apply it so that we can achieve better computational results.

{{} The organization of this paper {{} is as follows}. In next section a brief description of essential concepts of APT is reviewed. In Sec.\ref{pQCD} evolution of parton densities and nucleon structure function, using Jacobi transformation are discussed. Sec.\ref{fapt} is devoted to illustration of the structure function in MPT model. Based on this model the Gottfired sum rule is considered in Sec.\ref{sec:sumrule}. Finally summary and conclusion is presented in Sec.\ref{summery}}

\section{{ Basic concepts in analytic perturbation theory}}\label{FRACTIONAL}

 As we mentioned, underlying pQCD coupling suffers from unphysical {{} Landau singularities} at $Q^2 \sim \Lambda ^2$ . Therefore we can not apply it in the low momentum and that is a motivation to use other approaches, {{} especially analytic QCD (anQCD)} to achieve fairly accurate results for physical quantities.
 In this approach we have analytic couplings $A_{\nu}$ which are free from aforementioned problems. In  the following content we will represent the main elements of APT.
 Application of Cauchy theorem to the running coupling {{} $a(Q^2)$,  where $a(Q^2) = \alpha_s(Q^2)/\pi$,} gives us the following spectral relation in general anQCD \cite{Ayala:2014pha,Ayala_2015}:
  \begin{equation}\label{eq:1}
 A_{1}(Q^2) =  \frac{1}{\pi}\int_{\sigma = M_{th}^{2}}^{\infty}\frac{d\sigma \rho_{1}(\sigma)}{(\sigma + Q^2)}\;,
 \end{equation}
  where
 \begin{equation}\label{eq:2} {{}
\rho_{1}(\sigma) \equiv Im A_1(-\sigma - i\varepsilon)\;.}
\end{equation}
Different approaches to consider  the discontinuity function $\rho _{1}({\sigma})$ and the coupling function $A_{1}(Q^2)$ will lead to the various  anQCD models. As we pointed out before {{} $A_1(Q^2)$ is the anQCD-analog of the underlying pQCD coupling $a(Q^2) = \alpha_s(Q^2)/\pi$, i.e., at $\sigma \gg \Lambda$ we have ${\equiv Im} a(-\sigma - i \varepsilon) = {\equiv Im} A_1(-\sigma - i \varepsilon)$. Let us denote by $A_{\nu}$ the anQCD-analog of the pQCD power $a^{\nu}$ (where $\nu$ is not necessarily integer).} As an important point we should notice that there is not standard algebra for
$A_{\nu}$, i.e., $A_{\nu} A_{\mu} \not\equiv A_{\nu+\mu}$ or {{} $A_{\nu+\mu} \not\equiv (A_{\nu})^{\mu}$. For the construction of $A_{\nu}$ in a general anQCD, we follow Ref.~\cite{Cveti__2012}.}

Correct analogs $A_{n}(Q^2)$ of the powers $a^n (Q^2)$ will be achieved, using the logarithmic derivatives of {{} $A_1$} \cite{Ayala:2014pha}:
 \begin{equation}\label{eq:3}
 {\tA}_{n+1}\equiv \frac{(-1)^n}{\beta _{0}^{n}n!}(\frac{\partial}{\partial ln Q^2})^n A_{1}(Q^2)\;.
 \end{equation}
  It is obvious that with n=0 we get $\tA_{1}\equiv A_{1}$. Here $\beta_{0}=\frac{1}{4}(11-\frac{2}{3}N_{f}) $ is  the first coefficient of QCD-$\beta$ function which is scheme independent where this function is governing by Renormalization Group Equation (RGE) for QCD runnning coupling constant.
  Substituting $A_{1}$ in Eq.(\ref{eq:1}) into Eq.(\ref{eq:3}) will lead to
 \begin{equation}\label{eq:4}
 \tA_{n+1}(Q^2) = \frac{1}{\pi}\frac{(-1)}{\beta_{0}^{n}\Gamma(n+1)}\int_{0}^{\infty}\frac{d\sigma}{\sigma}\rho_{1}(\sigma)Li_{-n}(\frac{-\sigma}{Q^2})\;.
 \end{equation}
 In this equation  $n$ is an integer number and it  can be extended to noninteger index $\nu$ {{} as follows  \cite{Cveti__2012}:}
 \begin{equation}\label{eq:5}
  \tA_{\nu+1}(Q^2) = \frac{1}{\pi}\frac{(-1)}{\beta_{0}^{\nu}\Gamma(\nu +1)}\int_{0}^{\infty}\frac{d\sigma}{\sigma}\rho_{1}(\sigma)Li_{-\nu}(\frac{-\sigma}{Q^2})\;.
 \end{equation}
 Here $Li_{-\nu}(\frac{-\sigma}{Q^2})$ is polylogarithm function.  It should be  noted that the integral in Eq.(\ref{eq:5})  is converging at low   $\sigma$ for $\nu > -1$ where polyloghartitm function is approximated by $Li_{-\nu}(-z)\sim ln^{-\nu}z$.
 The analytic analogs $A_{\nu}(Q^2)$ can be constructed as linear combinations of $\tA_{\nu + m}$'s:
 \begin{equation}\label{eq:6}
 A_{\nu}(Q^2) = {\tA}_{\nu}+\sum_{{m\geq 1}} {\tk}_{m}(\nu){\tA}_{\nu +m}\;.
 \end{equation}
The coefficients ${\tk}_{m}(\nu)$  in  Eq.(\ref{eq:6}) have been  determined in \cite{Cveti__2012}.  Using the analytic coupling constant we can do the required calculations for quantities which contain  noninteger power expansion of the coupling constant.

The specific anQCD model that has been used in this paper, is named Massive Perturbation Theory (MPT) in which to achieve a holomorphic coupling,  an effective mass is attributed to gluon such as \cite{MPT,Ayala:2014pha}
\begin{equation}\label{eq:7}
A_{1}^{(MPT)}(Q^2,N_{f}) = a(Q^2 + m_{gl}^{2},N_{f})\;.
\end{equation}
The mass scale $m_{gl}\sim 0.5 - 1\; GeV$ refers to gluon
mass where $m_{gl}^{2} = 0.7 Gev^{2}$ is considered here. Since $m_{gl}^{2} > \Lambda ^{2}$, we get a coupling that is analytic {{} even to scales less than $\Lambda$.}
At high energy $A_{1}^{MPT}(Q^2)$ tends to the pQCD coupling $a(Q^2)$ .  It can be seen that the difference of MPT with respect to the  underlying pQCD coupling would be given by \cite{Ayala:2014pha}:
\begin{equation}\label{eq:8}
A_{1}^{MPT}(Q^2 , N_{f}) -a(Q^2,N_{f})\sim\frac{m_{gl}^{2}}{Q^{2}ln^{2}(\frac{Q^2}{\Lambda^2})}\;.
\end{equation}

In the following sections  we can observe the benefits of   MPT  model in comparing with pQCD approach. As an adjunct to this issue we plot in Fig.1 the running coupling constant in two MPT and FAPT models and compare it with the underlying pQCD coupling

\begin{figure}[!htb]
	\vspace*{0.5cm}
	\includegraphics[clip,width=0.5\textwidth]{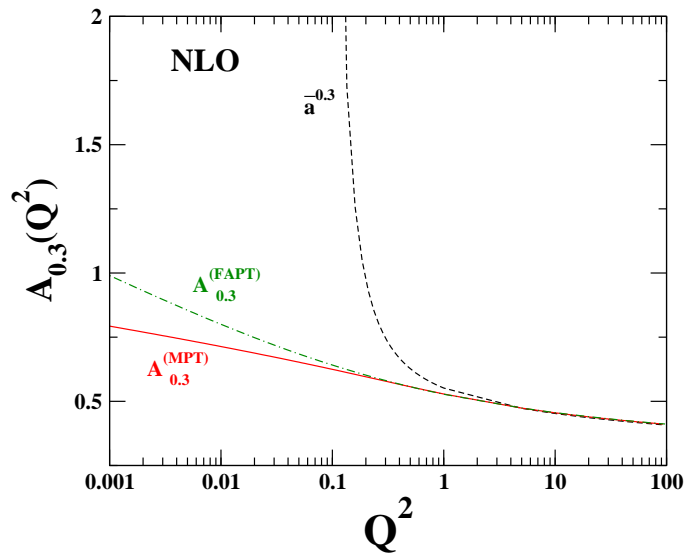}
\hspace{1cm}
	\begin{center}

\caption{(Color online) \small The coupling constants, using  two MPT and FAPT models with fraction index $\nu$ =0.3, are presenting. The moderate behaviour  of the running coupling constant in these models at low energy scales in comparison to underlying pQCD is obvious.}
	\end{center}\label{fig1}
\end{figure}

\section{Jacobi transformation and parton densities evolutions}\label{pQCD}
 To extract the unpolarized NSF in terms of $Q^2$ energy scale we need to do the energy evolution for both the  singlet and nonsinglet sectors of SF. Here  we start by singlet densities where splitting functions are governing their evolution. The singlet quark distribution of hadron is defined by

\begin{equation}\label{eq:9}
\Sigma(x,Q^2)=\sum_{i=1}^{N_{f}}[ q_{i}(x,Q^2)+\bar{q_{i}}(x,Q^2)]\;.
\end{equation}

Here $q_{i}(x,Q^2)$ and $\bar{q_{i}}(x,Q^2)$ represent the respective number densities of quarks and antiquarks as a function of the carried momentum fraction x. The subscript i indicates the flavor of the (anti)quark and $n_f$ stands for the the number of effectively massless flavours.
Suppressing the fractional dependencies, the coupled evolution equations for the singlet patron and gluon distributions read
\begin{equation}\label{eq:10}
\frac{d}{dlnQ^2}
\begin{pmatrix}
\Sigma\\
g
\end{pmatrix}
=
\begin{pmatrix}
P_{qq} & P_{qg}\\
P_{gq} & P_{gg}
\end{pmatrix}
\otimes
\begin{pmatrix}
\Sigma_{0}\\
g_{0}
\end{pmatrix}
,
\end{equation}\label{eq:11}
where $\otimes$ stands for convolution  integral in the momentum variable,
\begin{equation}
[a\otimes b](x)\equiv \int_{x}^{1}\frac{dy}{y}a(y)b(\frac{x}{y}).
\end{equation}
{{} The corresponding gluon distribution, $g(x,Q^2)$, is denoted briefly here by $g$.}

The quark-quark splitting function $P_{qq}~$\cite{Floratos:1981hs} in Eq.(\ref{eq:10}) can be expressed as \cite{Vogt:2004ns}
\begin{equation}\label{eq:12}
P_{qq}=P_{ns}^{+}+N_{f}(P_{\bar{q}q}^{s}+P_{{q}q}^{s})\equiv P_{ns}^{+}+P_{ps}\;.
\end{equation}
Here $P_{ns}^{+}$ is the non-singlet splitting function. The quantities $P_{qq}^{s}$ and $P_{\bar{q}q}^{s}$ are the flavour independent sea contributions to the quark-quark and quark-antiquark splitting functions respectively.
The gluon-quark entries in Eq.(\ref{eq:10}) are given by
\begin{equation}
P_{qg}=N_{f}P_{q_{i}g},\;\;
P_{gq}=P_{gq_{i}}\;.
\end{equation}
In terms of the flavour independent splitting functions one can write $P_{q_{i}g}=P_{\bar{q}g}$ and $P_{gq_{i}}=P_{g\bar{q}}$.

The required calculations can now be continued in Mellin-N space, using Mellin transformation:
\begin{equation}
a(N)=\int_{0}^{1}dxx^{N-1}a(x)\;.
\end{equation}
Then by transforming all needed quantities to {{} Mellin (moment) space, the solution of  Eq.(\ref{eq:10}) at  NLO accuracy is rendered by:}

\begin{eqnarray}\label{eq:13}
\begin{pmatrix}
\Sigma \\
g
\end{pmatrix}
 &=& \Bigg\{ \left( \frac{a_s}{a_0} \right)^{-r_-}
\bigg[\, \eV_- + (a_0 - a_s)\:  \eV_- \RV_1 \eV_- \nonumber \\
& &  \qquad \mbox{}
- \bigg(a_0 - a_s \left( \frac{a_s}{a_0} \right)^{r_--r_+} \bigg)
\,\frac{\eV_- \RV_1 \eV_+}{r_+-r_--1} \,\bigg]
\:\nonumber\\ &  &   + \: (\, + \leftrightarrow - \,) \Bigg\}
\begin{pmatrix}
\Sigma_{0}\\
g_{0}
\end{pmatrix}\;.
\end{eqnarray}
{{} Here we explicitly define  $a_s=\alpha_s/(4 \pi) = a/4$.}  In the last line  the following  recursive abbreviations have been used \cite{Vogt:2004ns}

\begin{eqnarray}
\RV_0 \:&\equiv&\: \frac{1}{\beta_0} \PV^{(0)} \:\: , \nonumber\\
\RV_k \:&\equiv&\: \frac{1}{\beta_0} \PV^{(k)} - \sum_{i=1}^{k}
b_i \,\RV_{k-i}\;,
\end{eqnarray}
with $b_k\equiv \beta_k /\beta_0$. Furthermore for the $r_{\pm}$ one can write
\begin{equation}
r_{\pm}=\frac{1}{2\beta _{0}}\left[P_{qq}^{0}+P_{gg}^{0}\pm \sqrt{(P_{qq}^{0}-P_{gg}^{0})^{2}-4P_{qg}^{0}P_{gq}^{0}} \right]\;,
\end{equation}
where the following relation for  $\eV_{\pm}$ is defined:
\begin{equation}
\eV_{\pm} = \frac{1}{r_{\pm}-r_{\mp}} \Big[ \RV_0-r_{\mp}\IV \Big]\;,
\end{equation}
in which $\IV$ is representing a unique $2\times2$ matrix.

{{} In non-singlet case in order to decouple the combination, {{} it is needed} to use the general structure of (anti-) quark (anti-) quark splitting functions  {{} as it follows}\cite{Vogt:2004ns}
\begin{equation}
\begin{aligned}\label{eq:27}
P_{q_{i}q_{k}} &= P_{\bar{q}_{i}\bar{q}_{k}}=\delta_{ik}P_{qq}^{V}+P_{qq}^{s}\;,\\
P_{q_{i}\bar{q}_{k}} &= P_{\bar{q}_{i}q_{k}}=\delta_{ik}P_{q\bar{q}}^{V}+P_{q\bar{q}}^{s}
\end{aligned}
\end{equation}
The flavour asymmetries $q_{ns}^{\pm}$ and the total valence distribution $q_{ns}^{V}$ and their corresponding splitting functions {{} are given by} \cite{Vogt:2004ns},
\begin{equation}
\begin{aligned}
q_{ns,ik}^{\pm} &= q_{i} \pm \bar{q}_{i}-(q_{k} \pm \bar{q}_{k})\;,\\
q_{ns}^{V} &= \sum_{r=1}^{n_{f}}(q_{r}-\bar{q}_{r})\;,\\
P_{ns}^{\pm} &= P_{qq}^{V} \pm P_{q\bar{q}}^{V}\;,\\
P_{ns}^{V} &= P_{qq}^{V}-P_{q\bar{q}}^{V}+n_{f}(P_{qq}^{s}-P_{q\bar{q}}^{s})\equiv P_{ns}^{-}+P_{ns}^{s}\;.
\end{aligned}
\end{equation}

For non-singlet quark distributions evolution  a similar process {{} exists like the} singlet case, but with the obvious simplification that no spurious complexity occur. Consequently the non-singlet evolution can be written as it follows \cite{Vogt:2004ns} :
\begin{eqnarray}\label{eq:22}
 \begin{aligned}
q^{\pm ,V}(a_{s})= {} & exp\Bigg [R_{1}^{\pm ,V}(a_{0}-a_{s})\Bigg]\Bigg(\frac{a_{s}}{a_{0}}\Bigg)^{-R_{0}^{\pm ,V}}\\
\times q^{\pm ,V}(a_{0})
\end{aligned}
\end{eqnarray}
In which $R_{0}^{\pm ,V}=\frac{1}{\beta _{0}}P^{(0)\pm ,V}${{} and $R_{1}^{\pm ,V}=\frac{1}{\beta_0}P^{(1)\pm ,V}-\beta_{1}R_{0}^{\pm ,V}$} are defined  based on non-singlet splitting functions {{} where $\beta_0$ and $\beta_1$ are the first  two universal coefficient of QCD $\beta$-function} .
Accordingly Eq.(\ref{eq:22}) at next-to-leading-order (NLO) accuracy can be written:

\begin{eqnarray}\label{eq:23}
 \begin{aligned}
q^{\pm ,V}_{NLO}(a_{s})= {} & \Bigg [1+R_{1}^{\pm, V} (a_{0}-a_{s})\Bigg ] (\frac{a_{s}}{a_{0}})^{-R_{0}^{\pm ,V}}\\
& \times q^{\pm ,V}(a_{0})\;.
\end{aligned}
\end{eqnarray}

}
Finally, using  Eq.(\ref{eq:13}) and Eq.(\ref{eq:23}) we can obtain the nucleon structure function at the NLO accuracy  in
Mellin (moment) N-space as it follows

\begin{eqnarray}\label{eq:14}
F_{2}(N,Q^2)&=&[C_{2q}^{(0)}(N)+a_{s}(Q^2)C_{2q}^{(1)}(N)]
\sum_{i=u,d,s}e_i^2q_{i}(N,Q^2)\nonumber\\
&&+a_{s}(Q^2)C_{2g}^{(1)}(N)\frac{1}{f}\sum_{i=u,d,s}e_i^2g(N,Q^2)\;.
\end{eqnarray}
Here $C^{k}(N)$s are Wilson coefficient functions which have been calculated in \cite{Vogt:2004mw}.
Using Jacobi transformation, as was mentioned before, is an adequate method to convert the calculated results from moment N-space  to Bjorken x-space. Details of this method has been described in \cite{Koekoek_2000}.
According to this method we can define the NSF, based on the following relation:

\begin{equation}\label{eq:15}
F_{2}(x,Q^2)=x^{\beta}(1-x)^{\alpha}\sum_{x=0}^{\infty}a_{n}^{\alpha ,\beta}(Q^2)\Theta_{n}^{(\alpha , \beta)}(x)\;.
\end{equation}
Here $a_{n}^{\alpha ,\beta}(Q^2)$ is an expansion coefficient and $\Theta_{n}^{(\alpha , \beta)}(x)$ is denoting to the Jacobi polynomials {{}and they are related to each other by:}

\begin{equation}\label{eq:16}
a_{n}^{\alpha ,\beta}(Q^2)=\int_{0}^{1}F_{2}(x,Q^2)\Theta_{n}^{(\alpha , \beta)}(x)dx\;.
\end{equation}
By sunstituting $\Theta_{n}^{(\alpha , \beta)}(x)= \sum_{k=0}^{n}C_{k,n}^{(\alpha ,\beta)}x^k$ in Eq.( \ref{eq:16}) one can get:

\begin{equation}\label{eq:17}
a_{n}^{\alpha ,\beta}(Q^2)=\int_{0}^{1}F_{2}(x,Q^2) \sum_{j=0}^{n}C_{j,n}^{(\alpha ,\beta)}x^j dx
\end{equation}

Putting Eq.(\ref{eq:17}) into  Eq.(\ref{eq:15}) and using the SF in moment-N space by$M_{F_{2}}(N,Q^2)=\int_{0}^{1}x^{j-2}F_{2}(x,Q^2)dx$  we can achive to SF in Bjorken x-pace as it follows~\cite{Kataev:1997nc}:

\begin{eqnarray}\label{eq:18}
F_{2}(x,Q^2)=x^{\beta}(1-x)^{\alpha}\sum _{n=0}^{N_{max}}\Theta _{n}^{(\alpha ,\beta)}(x)\nonumber\\
\sum_{j=0}^{n}C_{j,n}^{(\alpha ,\beta)}M_{F_{2}}(j+2,Q^2)
\end{eqnarray}

{{}\section{Unpolarized nucleon structure function and the MPT model}\label{fapt}}

{{} To do the required computations to extract the nucleon structure function we need first the parton distribution functions (PDFs) at initial  energy scale, $Q_0$, as the inputs. For this purpose the following parameterized functions are suggested  \cite{Jimenez_Delgado_2014}}.

\begin{equation}\label{eq:25}
\begin{gathered}
 xu_{v}(x,Q_{0}^{2})=N_{u}x^{a_{u}}(1-x)^{b_{u}}(1+A_{u}\sqrt{x}+B_{u}x+C_{u}x^{2}),\\
 xd_{v}(x,Q_{0}^{2})=N_{d}x^{a_{d}}(1-x)^{b_{d}}(1+A_{d}\sqrt{x}+B_{d}x+C_{d}x^{2}),\\
xg(x,Q_{0}^{2})=N_{g}x^{a_{g}}(1-x)^{b_{g}}(1+B_{g}x^{\alpha _{g}}(1-x)^{\beta _{g}}),\\
 x\Sigma(x,Q_{0}^{2})=N_{\Sigma}x^{a_{\Sigma}}(1-x)^{b_{\Sigma}}(1+A_{\Sigma}\sqrt{x}+B_{\Sigma}x),\\
  x\Delta(x,Q_{0}^{2})=N_{\Delta}x^{a_{\Delta}}(1-x)^{b_{\Delta}}(1+A_{\Delta}\sqrt{x}+B_{\Delta}x),\\
  xs(x,Q_{0}^{2})=N_{s}x^{a_{s}}(1-x)^{b_{s}}(1+A_{s}\sqrt{x}+B_{s}x)
\end{gathered}
\end{equation}

In Eq.(\ref{eq:25}) {{}  these definitions are used:} $u_{v}=u-\bar{u}$, $d_{v}=d-\bar{d}$, $\Sigma =\bar{u}+\bar{d}$ and $\Delta = \bar{d}-\bar{u}$. {{} All unknown parameters, including the normalization factors are obtained via the fitting over the related data \cite{Jimenez_Delgado_2014}. The results are listed in Table.\ref{table1}}.

\begin{table}
	\begin{center}
	\begin{tabular}{cccccc}
	\hline
	\multicolumn{6}{c}{NLO} \\ \hline
	N$_{u}$ & 1.63 & N$_{d}$ & 7.4 & N$_{g}$ & 2.95 \\
	a$_{u}$ & 0.55 & a$_{d}$ & 0.92 & a$_{g}$ & 0.047 \\
	b$_{u}$ & 3.61 & b$_{d}$ & 4.6 & b$_{g}$ & 6.1 \\
	A$_{u}$ & 0.8 & A$_{d}$ & -2.8 & B$_{g}$ & 0 \\
	B$_{u}$ & 4.7 & B$_{d}$ & 4.5 & $\alpha _{g}$ & 0 \\
	C$_{u}$ & -0.1 & C$_{d}$ & -2 & $\beta _{g}$ & 0 \\ \hline
	N$_{\Sigma }$ & 0.164 & N$_{\Delta }$ & 57 & N$_{s}$ & 0.03 \\
	a$_{\Sigma }$ & -0.19 & a$_{\Delta }$ & 2.29 & a$_{s}$ & -0.28 \\
	b$_{\Sigma }$ & 8.42 & b$_{\Delta }$ & 18.6 & b$_{s}$ & 8.42 \\
	A$_{\Sigma }$ & 1.9 & A$_{\Delta }$ & 1 & A$_{s}$ & 1.9 \\
	B$_{\Sigma }$ & 10 & B$_{\Delta }$ & 0 & B$_{s}$ & 10 \\ \hline
\end{tabular}
\caption{\small Numerical values of the free parameters in Eq.~(\ref{eq:25}) at $Q_0^2=2\;GeV^2$ in the NLO accuracy ~\cite{Jimenez_Delgado_2014}.}
\end{center}\label{table1}
\end{table}

The computations of this paper is done in mathematica environment, using anQCD.m package \cite{Ayala:2014pha} and we are going to calculate analytic coupling constant corresponding to the underlying pQCD coupling, {{} realising the presented powers} in Eq.(\ref{eq:14}).
The relevant mathematica command of MPT coupling constant is $AMPTNl[N_{f},\nu,Q^{2},m_{gl}^{2},\Lambda^{2}]$ which returns the N-loop ($N=1,2,3,4$) analytic MPT coupling $A_{\nu}^{(MPT,N)}(Q^{2},m_{gl}^{2},N_{f})$, {{} including the fractional index $\nu$} at fixed number of active quark flavours $N_{f}$, with $Q^2$ in the Euclidean domain ($Q^{2}>0$). In as much we do calculations at NLO approximation, {{} the $N$-loop is fixed at 2 (i.e., we use 2-loop MPT)}. The other commands of various anQCD models, for analytic coupling constant achievement, have been described in \cite{Ayala:2014pha}. Interested reader is encouraged to read also  \cite{Cveti__2012}.
{{} For simplicity we use  the the notation $a_{s}^{\nu}\equiv\frac{a^{\nu}}{4^{\nu}}$ and subsequently $A_{s,\nu}\equiv\frac{A_{\nu}}{4^{\nu}}$,} so the mentioned command becomes  $\frac{AMPT2l[3,\nu,Q^{2},m_{gl}^2,\Lambda^{2}]}{4^{\nu}}$. Here $N_{f}=3$, $m_{gl}^{2}=0.7$ $GeV^2$ and $\Lambda^2=0.35\;  GeV^2$.

{{} To employ the MPT model to extract the nucleon structure function, one may do it by applying the model to the evolution equations for singlet and non-singlet sectors of parton densities, given by  Eq.(\ref{eq:13}) and Eq.(\ref{eq:23}) and finally using the MPT  model separately to Eq.(\ref{eq:14}), containing the Wilson coefficients. This is not admissible since Wilson coefficients  and parton densities are not directly observable and it is  the nucleon structure function that should be analyzed, and not the factors separately. On this based we need to resort to Eq.(\ref{eq:14}) and employ the MPT model entirely on it. Hence each analytical coupling constant is utilized where part of its total  exponent number is coming from the evolved parton densities and the rest is back  to exponent of coupling constant behind Wilson factors. This procedure is completely corresponding to the property of analytical couplings, presented before by $A_{\nu}A_{\mu} \neq A_{\nu+\mu}$ or $A_{\nu+\mu} \neq (A_{\nu})^ {\mu}$.

In fact what we need lastly  to calculate can be given summarily by:

\begin{eqnarray}\label{eq:144}
&&F_{2}(N,A_{\nu+1}(Q^2))=[C_{2q}^{(0)}(N)+A_{1}(Q^2)C_{2q}^{(1)}(N)]\nonumber\\
&&\sum_{i=u,d,s}e_i^2q_{i}(N,A_{\nu}(Q^2))+A_{1}(Q^2)C_{2g}^{(1)}(N)\times\nonumber\\
&&\frac{1}{f}\sum_{i=u,d,s}e_i^2g(N,A_{\nu}(Q^2))\;,
\end{eqnarray}
{{} where we replace $A_1 A_{\nu} \mapsto A_{\nu+1}$ ($ \not= A_1 A_{\nu}$).}

Considering the numerical values for the required paremeters in analytical coupling, the utilized Mathematica command for  $A_{\nu}(Q^2)$ coupling would be $AMPT2l[3,\nu,Q^{2}, 0.7,0.35]$ where  $\nu$ index is determined via the evolution processes for singlet and  gluon densities and also non-singlet density. In practical calculations this index takes the following qualities: $\nu =-R_0, 1-R_0, 2-R_0, -r_{-},1-r_{-},2-r_{-},r_+,1-r_+,2-r_+$.}

Using available data at different energy scales, makes us the possibility to present  the $Q$ dependence of  $\alpha$ and $\beta$ Jocobbi parameters  in Eq.(\ref{eq:18}) as are following:

\begin{eqnarray}\label{eq:25-2}
	\alpha&=& -465.737 + 553.088\;\exp(Q^2)+\frac{336.376}{\log(Q^2)}\nonumber\\
	\beta&=&11.158 +\frac{7.670 Q^2}{\log(Q^2)}+14.308\;\sqrt{Q^2}\;\log(Q^2) \nonumber\\
\end{eqnarray}

\begin{figure}[!htb]
	\vspace*{0.5cm}
	\includegraphics[clip,width=0.5\textwidth]{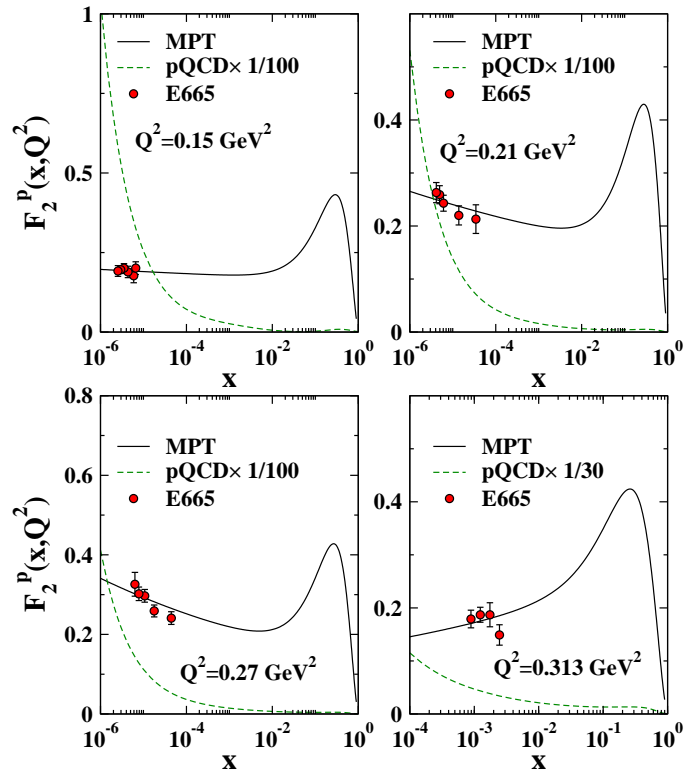}
	\begin{center}
		\caption{(Color online) \small Unpolarized nucleon structure function, $F_{2}^p(x,Q^2)$, as a function of $x$-Bjorken variable, using MPT  model (solid line) and underlying pQCD (dashed line). Comparison with the available experimental data \cite{E665:1996mob}} has also been done.
	\label{fig2}
	\end{center}

\end{figure}

We depict in Fig.\ref{fig2} the  $F_{2}^p(x,Q^2)$ structure function verses $x$  Bjorken variable at different energy scales $Q^2=0.15,0.21,0.27$ and $0.313$ $GeV^2$, using MPT modle and compare them  with E665 experimental data \cite{E665:1996mob}. To indicate the adequate applicability of MPT model at low energy scales  we also add  to this figure  the results of underlying pQCD  for the $F_{2}^p(x,Q^2)$ structure function. To achieve more precise results, the computation of underlying  pQCD are done at the two loops approximation of coupling constant ~\cite{Cafarella:2005zj,vanNeerven:1999ca,vanNeerven:2000uj,JimenezDelgado:2008rdu}.

There is a few experimental data in mentioned energies but as it is shown in Fig.\ref{fig2}, an appropriate agreement  is standing between anQCD results and the available experimental data.

\section{ Gottfried sum rule in anQCD approach}\label{sec:sumrule}
{{} Since the advent of quark-parton model, sum rules for nucleon structure functions play an important role for establishing the model. One of the important sum rule is called Gottfried sum rule (GSR). Considering the isospin symmetry for parton densities in proton and neutron, the numerical value for GSR would be different from the reported value by NMC group \cite{Amaudruz:1991at} where they measured the electromagnetic structure function of nucleon through the deep inelastic scattring  of  muons from proton and deuterons.}
 {{} Here we take into account the  GSR such as to include its numerical values at low energy scales. It is then admissible to follow the related calculation, using MPT model.} As we referred above, this sum rule provides determination of light flavour asymmetry of the nucleon sea  and is given by \cite{Broadhurst_2004}:

 \begin{equation}\label{eq:19}
 \begin{aligned}
 S_{G} \equiv {} &  \int_{0}^{1}\frac{dx}{x}[F_{2}^{p}(x,Q^{2})-F_{2}^{n}(x,Q^{2})]\\
&  =\int_{0}^{1}\frac{1}{3}(u(x)-d(x)+\bar{u}(x)-\bar{d}(x))dx\\
& =\frac{1}{3}-\frac{2}{3}\int_{0}^{1}(\bar{d}(x)-\bar{u}(x))dx
 \end{aligned}
 \end{equation}

 The $S_{G}$ deviates from the expectation of simple quark model. In other words if the nucleon sea were flavour symmetric, i.e.  $\bar{u}(x)=\bar{d}(x)$, we then obtain the  GSR  to be $S_{G}=\frac{1}{3}$, but it is in contrast with NMC  collaboration data in lepton-nucleon DIS \cite{Amaudruz:1991at,Kabuss:1997sm,Arneodo:1994sh}. {{} Accordingly the following $S_G$ numerical value has been reported} \cite{Abbate:2005ct,Broadhurst_2004}:
 \begin{eqnarray}\label{20}
 S_{G}(Q^{2}=4\;GeV^2)=0.235\pm 0.026
 \end{eqnarray}
 This discrepancy can be associated with existence of perturbative effects in the nucleon sea, which generate light-quark flavour asymmetry $\bar{u}(x,Q^2)<\bar{d}(x,Q^2)$ over significant range of Bjorken variable x \cite{Broadhurst_2004}.
 For numerical values of GSR at some  specific energies, we can refer to \cite{Abbate:2005ct}.

 We apply the MPT model to calculate {{} GSR at low energy scale less than QCD cutoff, $\Lambda$, that is about {{}$0.35\;GeV^2$}.  The  arisen numerical results are listed in Table.\ref{table1}. To avoid from numerical difficulty, we take the low  limit of integration of $S_G$ in Eq.(\ref{eq:19}) to be   $10^{-7}$. Due to {{}nonexistence} of the gluon radiation at low energy scale, the  probability of  sea quark appearance is very low  and it is expected that the $S_G$ value approaches to $\frac{1}{3}$. Consideration of $S_G$ values at low energy scales, as listed in Table.{\ref{gsr}} confirms this reality  }

 \begin{table}[h!]
 \begin{center}
  \caption{\small {{} Theoretical} GSR values,using MPT model at various $Q^2$}
   \label{table1}
   \begin{tabular}{l|c}
   	\hline\hline
   \textbf{Q$^2$}GeV$^2$ & \textbf{S$_{\textbf G}$}\\
   \hline
   0.15 & 0.325\\
   0.21 & 0.312\\
   0.27 & 0.301\\
   0.313 & 0.294\\
   4     & 0.196\\
   \hline
    \end{tabular}\label{gsr}
 \end{center}
 \end{table}

\section{SUMMARY and Conclusion}\label{summery}
{{} Considering the nonexistence of the pQCD coupling at low spacelike momenta $0 < Q^2 \lesssim \Lambda$,} we employed an approach, called anQCD for the purpose of reforming and modifying the calculations at energy scale $Q^{2}<\Lambda^{2}$ to {{} evaluate} unpolarized nucleon structure {{} function at the mentioned momenta.}
In this way considering the importance of gluon density in the singlet sector of nucleon structure function computations, {{} we applied  anQCD approach, based on MPT model,} {{} which contains a gluon mass parameter.
Using this approach, specifically the MPT model, NSF is calculable at all energy scales $Q^2 > 0$} where at moderate and high energies MPT results for NSF are matched {{} to those of the underlying pQCD}. It is seen that at low energies $F_{2}(x,Q^2)$ behaviour is smoother {{}{{} than  in the  underlying pQCD.}}
 We encounter these facts in Fig.\ref{fig2} at $Q^{2}=0.15,0.21,0.27$ and $0.313\; GeV^{2}$ respectively. Consequently with due attention to the acceptable conformity between MPT results and available data, we conclude that results of anQCD approach, using MPT model, are more {{} reliable than those of the (underlying) pQCD at low energies.

 Also, we evaluated} Gottfried sum rule while  a nucleon sea flavour asymmetry ($\bar{u}(x,Q^2)<\bar{d}(x,Q^2)$) is considered. {{} The naive GSR indicates a difference in the value with respect to the experimental data, because according to the naive parton model for GSR, $S_G = 1/3$.} But experimental data shows {{} a deviation from $\frac{1}{3}$.} By applying anQCD approach, {{} specifically} MPT model, {{} we achieved a result closer to the experimental data.} {{}In addition to experimental energy scale, $Q^2=4\; GeV^2$,  we employed this model at energy scales $Q^{2}=0.15,0.21,0.27$ and $0.313Gev^{2}$  {{} due to the} applicability of this approach at low energies. Numerical results for $S_G$ at low energy scales, based on the MPT model, {{} gives the results in agreement with the behaviour of the parton densities, which is the correct behaviour at these scales.}

 The anQCD approach can be employed to calculate the nuclear structure function like $^3{He}$ and $^7{Li}$ with data multiplicity  for them at low energies. We hope to report on this issue as our further research task.}

\section*{ACKNOWLEDGMENTS}

S. A. T. is grateful to the School of Particles and Accelerators, Institute for Research in Fundamental
Sciences (IPM) to make the required facilities to do this project. The rest of authors are thankful the Yazd university 
to provide the warm hospitality in connection to this research project.

\end{document}